\documentclass[epjCONF, onecolumn]{svjour}
\usepackage{graphics}
\usepackage[square, sort, numbers]{natbib}

\usepackage[varg]{txfonts} 
\newcommand\apjl{{ApJ}}%
\newcommand\apjs{{ApJS}}%
\newcommand\aap{{A\&A}}%
\def\simgt{\lower.5ex\hbox{$\; \buildrel > \over \sim \;$}}
\def\simlt{\lower.5ex\hbox{$\; \buildrel < \over \sim \;$}}

\def\msun{M$_\odot$}
\def\Teff{$T_{\rm eff}$}

\def\alfa2d{MLT--$\,\alpha^{\rm 2D}$}

\def\Dnu{$\Delta\nu$}
\def\numax{$\nu_{\rm max}$}

\def\kepler{\mbox{\textit{Kepler}}}

\usepackage{graphicx}
\usepackage{amssymb}
\usepackage{journals}

\RequirePackage{lineno}

\session-title{Conference Title, to be filled}
\begin{document}
\title{Differential population studies using asteroseismology: solar-like oscillating giants in CoRoT fields LRc01 and LRa01}
\author{A. Miglio\inst{1}\fnmsep\thanks{\email{a.miglio@bham.ac.uk}} \and C. Chiappini\inst{2} \and T. Morel\inst{3} \and M. Barbieri\inst{4} \and W. J. Chaplin\inst{1} \and L. Girardi\inst{5} \and J. Montalb{\'a}n\inst{3} \and A. Noels\inst{3} \and M. Valentini\inst{3} \and B. Mosser\inst{6} \and F. Baudin\inst{7} \and L. Casagrande\inst{8} \and L. Fossati\inst{9} \and V. Silva Aguirre\inst{10} \and A. Baglin\inst{6}}


\institute{School of Physics and Astronomy, University of Birmingham, United Kingdom \and
Leibniz-Institut fur Astrophysik Potsdam (AIP),  Potsdam, Germany \and
Institut d'Astrophysique et de G\'eophysique de l'Universit\'e de Li\`ege, Li\`ege, Belgium \and
Dipartimento di Astronomia, Universit\`a di Padova, Padova, Italy \and
INAF-Osservatorio Astronomico di Padova,  Padova, Italy \and
LESIA, CNRS, Universit{\'e} Pierre et Marie Curie, Universit{\'e} Denis Diderot, Observatoire de Paris, France\and
Institut d'Astrophysique Spatiale, Universit{\'e} Paris-Sud,  France\and
Research School of Astronomy and Astrophysics, Mount Stromlo Observatory, The Australian National University, ACT 2611, Australia\and
Argelander Institut f\"ur Astronomie der Universit\"at Bonn, Bonn, Germany\and
Stellar Astrophysics Centre, Department of Physics and Astronomy, Aarhus University,  Denmark}

\abstract{Solar-like oscillating giants observed by the space-borne satellites CoRoT and \kepler\ can be used as key tracers of stellar populations in the Milky Way. When combined with additional photometric/spectroscopic constraints, the pulsation spectra of solar-like oscillating giant stars not only reveal their radii, and hence distances, but also provide well-constrained estimates of their masses, which can be used as proxies for the ages of these evolved stars.
In this contribution we provide supplementary material to the comparison we presented in Miglio et al. (2013) between populations of giants observed  by CoRoT in the fields designated LRc01 and LRa01.}
%
\maketitle

This contribution is organised as follows. In \S \ref{sec:dir}, we briefly recall how global stellar properties (mass, radius, distance) are determined using average seismic constraints, while we refer to  \cite{Miglio2013a} for a general motivation behind this study, the methods adopted, and main results obtained.
In \S \ref{sec:data} we present the data used in this study, along with independent verifications of the asteroseismic radius and mass determination available in the literature (\S \ref{sec:test}).  Comparisons between the properties of the population observed in LRa01 and LRc01 are presented in \S \ref{sec:comp}. Finally, a comparison of observed and synthetic populations computed using parameterised models of the Milky Way is discussed in \S \ref{sec:synthetic}.

\section{Determination of global stellar properties}
\label{sec:dir}
Radii and masses of solar-like oscillating stars can be estimated from the average seismic parameters that characterise their oscillation spectra: the so-called average large frequency separation (\Dnu), and the frequency corresponding to the maximum observed oscillation power (\numax).
We refer to \citep{Miglio2013a} for a discussion on the method.

We determine stellar radii and masses by combining the available seismic parameters $\nu_{\rm max}$ and $\Delta\nu$ with effective temperatures $T_{\rm eff}$. The latter are determined using 2MASS \citep{Skrutskie2006} $J$ and $K_s$ photometry and the colour-$T_{\rm eff}$ calibrations by \citep{Alonso1999}, which depend only weakly on metallicity.  2MASS colours were transformed into the CIT photometric system used by \citep{Alonso1999} using the relations available in \citep{Alonso1998} and \citep{Carpenter2001}.

We then compute luminosities  $L$ using the Stefan-Boltzmann law: $L=4\pi  R^2 \sigma T_{\rm eff}^4$, where $\sigma$ is the Stefan-Boltzmann constant.  The distance modulus of each star is determined as $K'_{s\rm0}-K_{s\rm0}$, where $K_{\rm s0}$ is the de-reddened apparent 2MASS $K_s$ magnitude, and $K'_{s\rm0}$ the absolute $K_s$ magnitude. The latter is obtained combining $L$ and the bolometric corrections from \citep{Girardi05}, which are based on Castelli \& Kurucz 2005 ATLAS9 model atmospheres, for the range of $T_{\rm eff}$ and $\log{g}$ into consideration.

We take into account the effect of interstellar extinction on the magnitude and colour of each star using the 3D model of Galactic extinction in the $V$ band ($A_{\rm V}$) by \citep{Drimmel2003}.  The extinction in the $J$ and $K_{\rm s}$ bandpasses were determined following \citep{Fiorucci2003} assuming the spectral energy distribution of a K1 giant.
Since the extinction is distance dependent, we iterate the procedure until the derived distance does not vary by more than $1\,\%$.   We choose to consider magnitudes in the near-IR  to reduce the effect of interstellar reddening in both the determination of $T_{\rm eff}$ and apparent de-reddened magnitudes.

\section{Data available and uncertainties on stellar properties}
\label{sec:data}
The CoRoT photometric time series used in this work were obtained in the
so-called exofield during the first long CoRoT runs in the direction of
the Galactic centre (LRc01) and in the opposite direction (LRa01). These
long runs lasted approximately 140 days, providing us with a frequency
resolution of about 0.08 $\mu$Hz. A first analysis of these data was made
by \citep{DeRidder09}, \citep{Hekker09} and \citep{Kallinger2010}.
Solar-like oscillations were detected in  435 and 1626 giants belonging to LRa01 and LRc01, respectively. 
To derive $M$,$R$, and distance we use  the values (and uncertainties) of \numax\ and \Dnu\ as determined by \citep{Mosser2010}. The typical uncertainty on \numax\ and \Dnu\ for the 150-d long CoRoT observations is of the order of  2.4\,\% and 0.6\,\%, respectively.   

The uncertainties on the stellar properties are estimated using Monte Carlo simulations adopting the following constraints:
\begin{itemize}
\item Apparent magnitudes: uncertainties on $J$ and $K_s$ are taken from the 2MASS photometry available in the EXODAT \citep{Deleuil2009} catalogue (the median uncertainty is 0.02 mag). 
\item Extinction/reddening: a random error in $A_V$ of 0.3 mag was considered both in determining \Teff\ photometrically and in de-reddening $K_s$ apparent magnitudes.
\item \Teff: for each star we considered two sources of uncertainty. The first (100 K) due to uncertainties on the colour-\Teff\ calibration itself   \citep{Alonso1999}. The second one due to uncertainties on reddening. This results in a median combined uncertainty on \Teff\ of $\sim 190$ K (calibration+reddening).
\end{itemize}

\begin{figure}
\centering
      \includegraphics[width=.95\hsize]{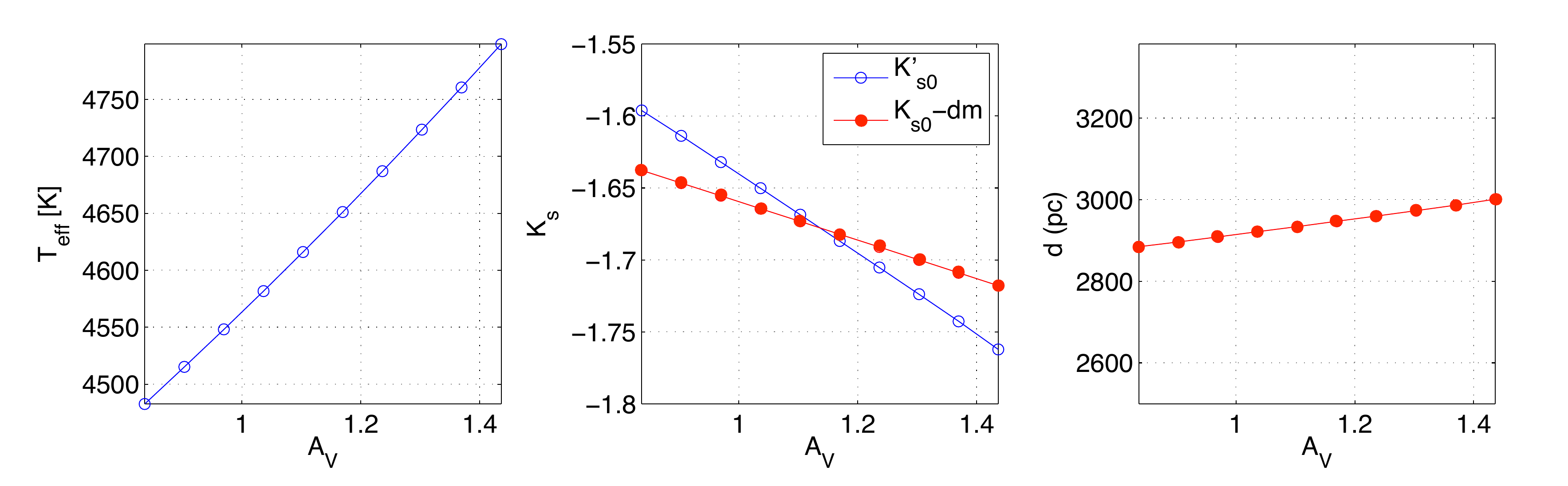}
      \caption{Example of the effect of reddening on the distance determination in a LRc01 star. {\it Left panel:} change in \Teff\ due to a change in the extinction $A_{\rm V}$. {\it Middle panel:} Absolute magnitude $K_{s0}'$ as a function of  $A_{\rm V}$ (blue line with open circles), and de-reddened apparent magnitude $K_{s0}$ (red line with filled circles). The distance modulus determined for the central value of $A_{\rm V}=0.17$ was subtracted from $K_{s0}$. {\it Right panel:} Distance as a function of the assumed $A_{\rm V}$.}
         \label{fig:redd}
\end{figure}

We notice that, when determining the distance, the effect of reddening affects not only  the de-reddened apparent  bolometric magnitude, but also the determination of $L$ (mostly via \Teff). A higher $A_V$ increases the estimated \Teff, hence $L$, but it also increases the apparent de-reddened luminosity $l$. Since $d \propto (L/l)^{1/2}$, the overall effect of reddening on the distance itself is  partly reduced (see Fig \ref{fig:redd}).  

By performing 500 realisations of the data assuming the uncertainties described above to be Gaussian,  we find a median intrinsic uncertainty of 5\,\% in distance, 3.6\,\% in radius, and 10\,\% in mass. 

To explore the effect of possible systematic offsets in the \Teff\ scale, we increased/decreased  \Teff\  by 100 K, and found that the distance estimate is affected by $\sim~2.5\,\%$ (the radius by 1\,\% and the mass by 3\,\%). 
These uncertainties may be further reduced when spectroscopic constraints will be available, however, since the estimates of $R$ and $M$ are based upon seismic scaling relations, care has to be taken on possible systematic uncertainties which are not taken into account in the estimates presented so far. 
Scaling relations for \Dnu\ and \numax\ are based on simplifying assumptions that need to be independently verified, and supported with empirical tests.

\subsection{Independent verification of the asteroseismic radius/distance and mass determination}
\label{sec:test}
In addition to preliminary investigations with stellar models (see e.g. \citep{Stello2009a, White2011}, the Supporting Online Material of \citep{Chaplin2011}, \citep{Miglio2012}), empirical tests of the asteroseismic radius determination are now available for: 

\begin{itemize}
\item 27 nearby dwarf and giant stars with precise Hipparcos parallaxes and asteroseismic constraints obtained with CoRoT data or ground-based spectroscopic campaigns. As shown in \citep{Miglio2012} for this set of targets the mean difference between seismic radii and those estimated by classical constraints is $-1.5\,\%$, with a standard deviation of $6\,\%$.
\item 22 dwarfs with available \kepler\ asteroseismic data and precise Hipparcos parallaxes (see \citep{SilvaAguirre2012}). The comparison shows that there is an overall agreement within $\pm\,5\%$  between the distance  obtained using the asteroseismic radii and  the Hipparcos one. 
Additional stringent tests were carried out by \citep{Huber2012}, including interferometric constraints on angular radii, showing that estimates of asteroseismic radii for main-sequence stars are accurate to better than 5\,\%. 
\item Giants belonging to the old-open cluster NGC6791. There is a remarkable agreement between the distance determination based on eclipsing binaries \citep{Brogaard2011}  and the one based on the radius determination of RGB stars (see \citep{Basu2011, Miglio2012}). While the seismic and ``classical''\footnote{Determined using distance modulus derived from eclipsing binaries, apparent magnitudes, bolometric corrections and $T_{\rm eff}$} radii agree within 5\,\%,  evidence for a small ($\sim 5\,\%$) but significant  discrepancy between the two determinations of radii was found for red-clump stars. 
\item Detailed modelling of individual frequencies available so far for two RGB giants \citep{DiMauro2011, Jiang2011} leads to radii and masses which agree within $1 \sigma$ (i.e. $\sim 2\,\%$ and $\sim 5\,\%$, respectively) with those derived combining \numax, \Dnu, and \Teff.
\item The radius distribution of stars in the observed CoRoT populations is dominated by an isolated peak at $\sim 11$ R$_\odot$ which is in agreement with the theoretically predicted radius of red-clump stars. Moreover, as  reported \citep{Miglio2012b}, the peak in the distribution of the absolute $K_s$ magnitudes of pulsating giants in the CoRoT fields is in agreement with the mean absolute $K_s$ magnitude of Hipparcos  red-clump giants \citep{Groenewegen08, Laney2012}. This agreement represents a further argument in support of the seismically derived distances, since in old composite populations the absolute $K_s$ magnitude of clump stars is expected to depend little on the details of the population itself \citep[see][]{Girardi01}. 
\end{itemize}
Based on the tests reported above, we consider a 5\,\% systematic uncertainty on the radius determination. This leads to an overall  uncertainty on the distance of  5\,\% (random) + 5\,\% (systematic due to radius determination via scaling relations) + 2.5\,\% (systematic due to the \Teff\ scale).

Independent tests of the mass determined using seismic scaling relations are less numerous and stringent: precise estimates of stellar mass are typically available only for visual/eclipsing binaries or via model-dependent estimates (e.g. isochrone fitting in clusters):
\begin{itemize}
\item Direct mass measurements are available only for a handful of dwarfs with detected solar-like oscillations \citep[see][]{Bedding2011b, Miglio2012c}. Self-consistency checks of the mass determination were performed in \citep{Miglio2012c} combining \numax\ or \Dnu\ alone with independent radii determinations, but given the large uncertainties in the data used, these comparisons could not test the consistency of the mass determination (using \Dnu\ or \numax) to better than 15-20\,\%. 

\item  As recalled above,  \citep{DiMauro2011} and \citep{Jiang2011} showed that, in the two cases analysed, detailed modelling of individual frequencies in (low-luminosity) giants leads to masses which are within 5\,\% those determined using scaling relations.  These tests should be performed for a much wider set of targets, encompassing stars of different mass, metallicity, and evolutionary state.

\item The strongest constraints are provided by giants in clusters.  Independent radius estimates of stars in NGC6791 were used in \citep{Miglio2012} to check consistency of the mass determination using the scaling relation for \numax\ or \Dnu\ alone. While no significant systematic effect was found on the RGB,  \citep{Miglio2012} suggested  that  a  relative correction to the \Dnu\ scaling relation should be considered between RC and RGB stars, affecting the mass determination of clump stars by $\sim 10\,\%$. 

By combining constraints from near-turnoff eclipsing binaries and stellar models, \citep{Brogaard2012} determined precisely the mass of stars on the lower RGB of NGC6719. Their value agrees within $\sim7\,\%$  with the average mass of RGB stars determined by \citep{Basu2011} and \citep{Miglio2012}.  The difference is however significant given the quoted uncertainties, and further work is needed to understand the origin of the discrepancy. 
\item  Finally, the mass of giants in NGC6819 derived from seismology \citep{Basu2011,  Miglio2012} is also in good agreement ($\sim 10\,\%$) with the one estimated from isochrone fitting \citep{Kalirai2004, Hole2009}, but the uncertainties and model-dependence of the latter method do not provide a stringent test to the seismically determined mass. 

\end{itemize}

\begin{figure}[ht]
\vspace{-1cm}
\centering
\includegraphics[width=.8\hsize]{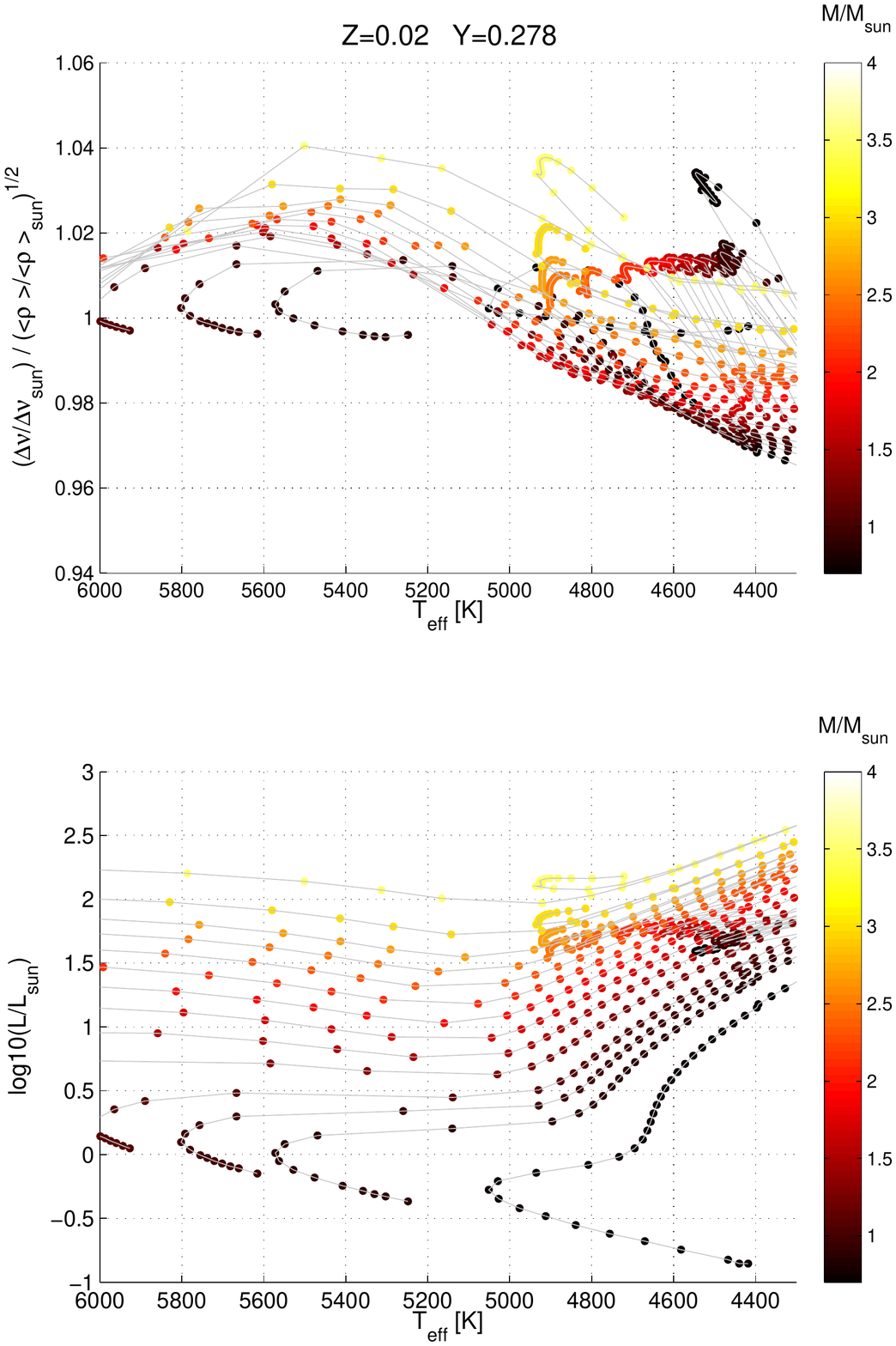}
\caption{{\it Upper panel:} Ratio $\Delta\nu/\Delta\nu_\odot$ to $(\rho/\rho_\odot)^{1/2}$ as a function of $T_{\rm eff}$ for models computed using the ATON code \citep{Ventura2008},  with initial heavy-elements mass fraction $Z=0.02$, initial helium mass fraction $Y=0.278$,  and mass between 0.7 \msun\ and 3.5 \msun. Adiabatic oscillation frequencies were computed using LOSC \cite{Scuflaire2008b} (see \cite{Montalban2010, Montalban2013} for more details about the models and the computation of oscillation frequencies). $\langle\Delta\nu\rangle$ is a gaussian-weighted average of the large frequency separation of radial modes (using a Gaussian of width \numax/2 centred in \numax). 
{\it Lower panel:} HR diagram showing evolutionary tracks corresponding to the models shown in the upper panel.}
\label{lab:scalings}
\end{figure}

It is clear that additional tests against independent estimates of mass and radius are needed, and in particular for metal-poor stars, which are under-represented in the above-mentioned tests. 
In addition to these empirical tests,  \Dnu\ computed from model frequencies may be used to correct, or at least investigate, possible shortcomings in the \Dnu\ scaling relation. Work in this direction was already presented in \cite{White2011} and extended, although for a specific case only, to He-burning stars in \cite{Miglio2012}. This approach captures both the deviation from the asymptotic expression due to the varying radial order of the observed modes, as well as to the different sound-speed distribution, hence acoustic radius, of stars with same mean density. In Fig. \ref{lab:scalings} we present an example of such model-predicted correction, which extend tests presented in \cite{White2011} to stars in the the core-He burning phase.  More detailed results will be presented in a forthcoming work.

Because we have much less confidence in theoretical computations of $\nu_{\rm max}$ -- which
rely on the complicated excitation and damping processes -- than we do in theoretical predictions of the oscillation frequencies, model-computed $\nu_{\rm max}$ have so far not been used in stellar properties estimation. More work is needed to understand the observed $\nu_{\rm  max}$ scaling, and theoretical studies have made progress on the problem (e.g., see \cite{Belkacem2011}).

\section{Comparison between populations observed in LRa01 and in LRc01}
\label{sec:comp}
Before comparing populations of giants observed in the two fields, we check whether different biases were introduced in the target selection in LRa01 and LRc01.

\subsection{CoRoT target selection}
\begin{figure}[ht]
\vspace{-3cm}
\begin{minipage}[t]{1.0\hsize}
\centering
      \includegraphics[width=.54\hsize]{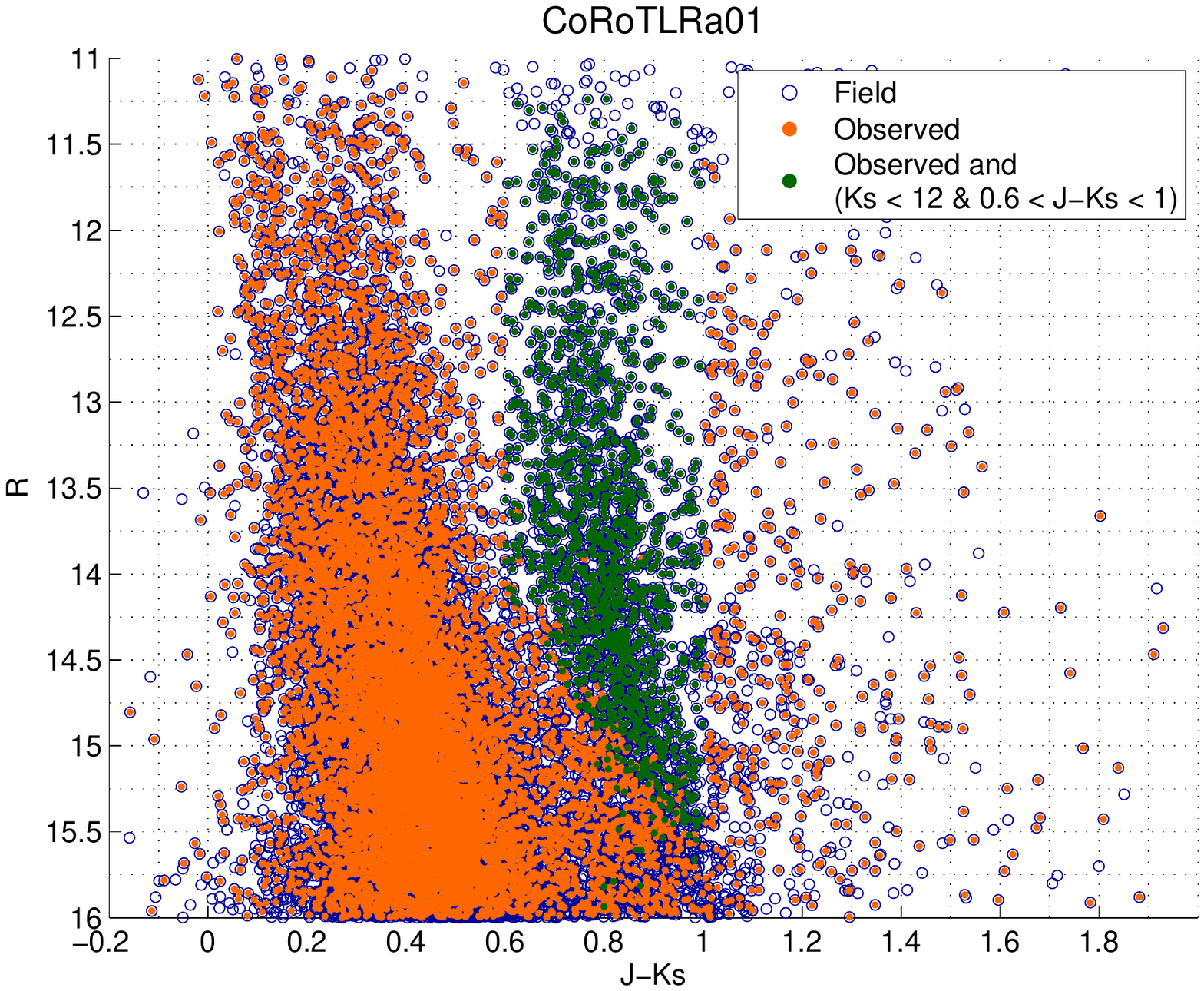}
      \hspace{-1.55cm}
    \includegraphics[width=.54\hsize]{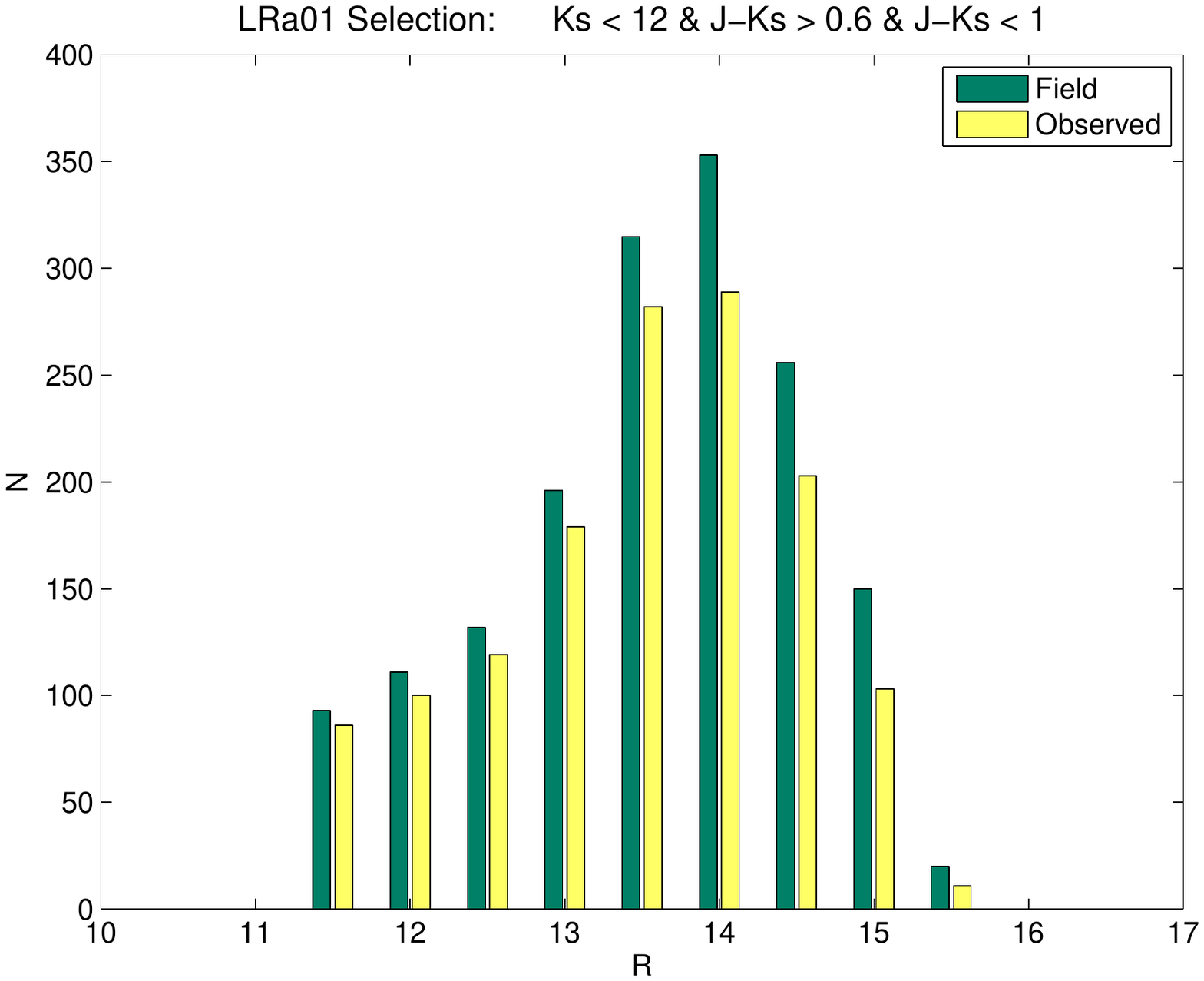}
\end{minipage}
\vspace{-5.5cm}

\begin{minipage}[t]{1.0\hsize}
\centering
      \includegraphics[width=.54\hsize]{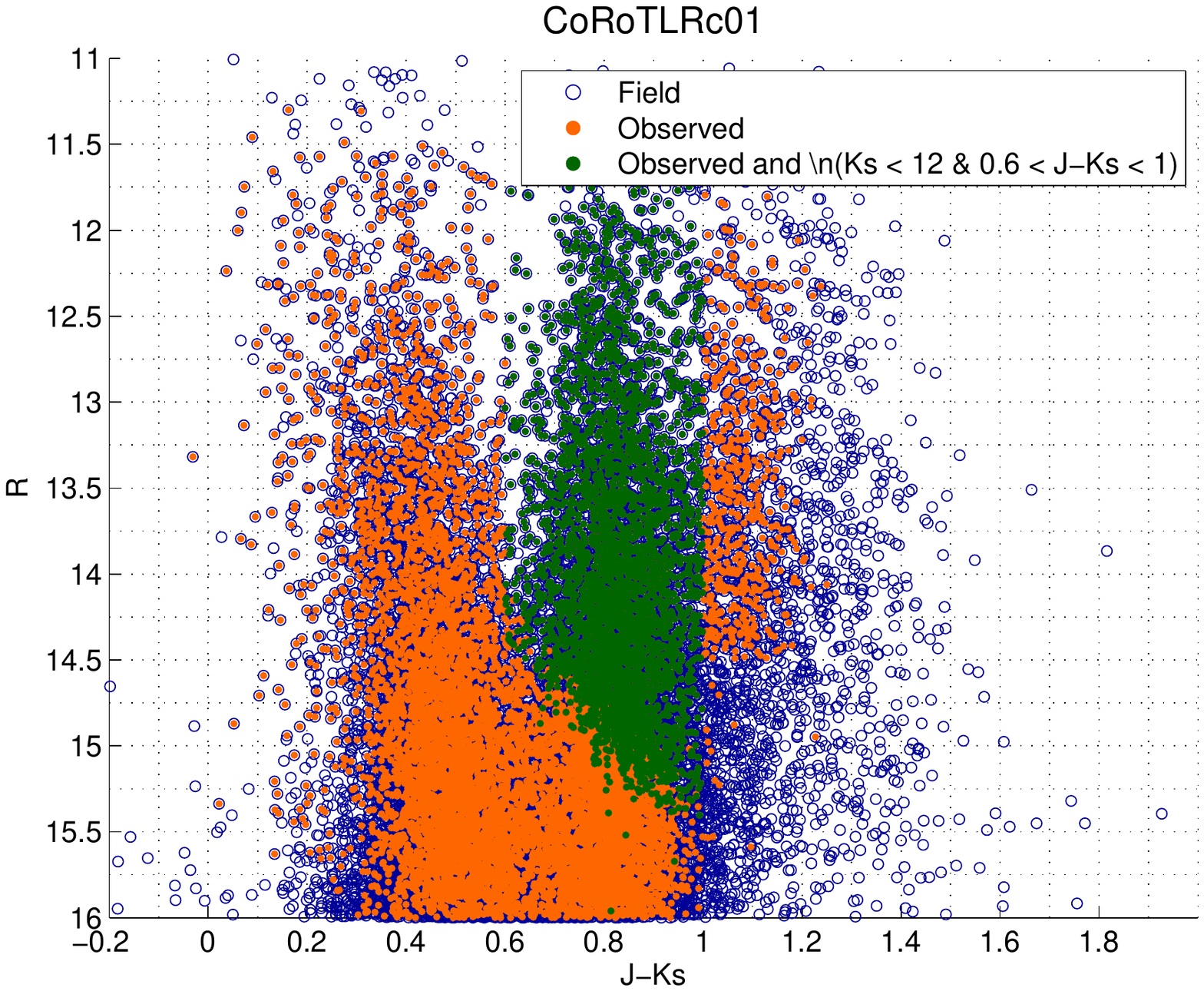}
      \hspace{-1.45cm}
    \includegraphics[width=.54\hsize]{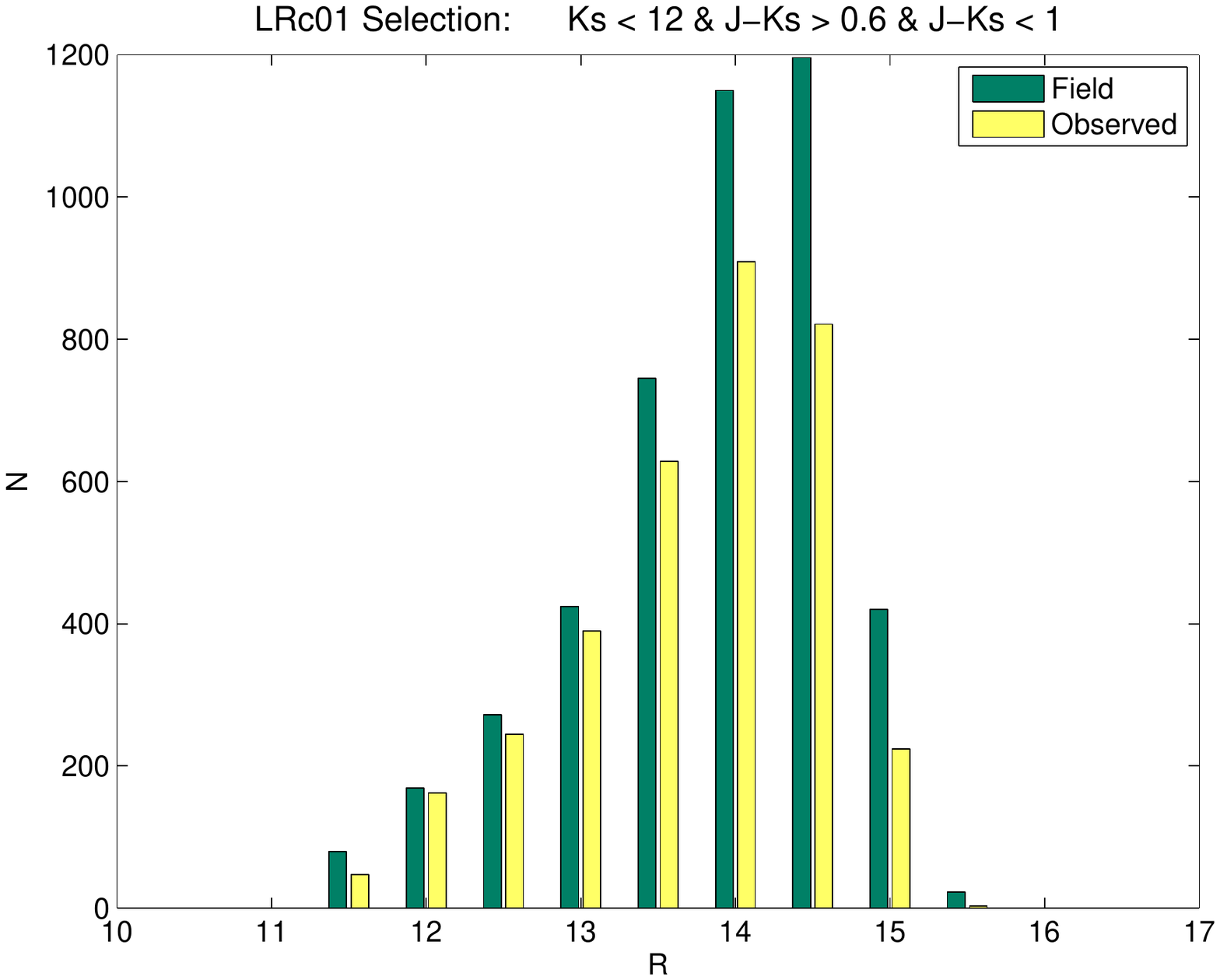}
\end{minipage}
\vspace{-2.5cm}

      \caption{{\it Upper-left  panel:} Colour-magnitude diagram of stars with $R < 16$ in the LRa01 field of view (blue circles). Stars observed by CoRoT (orange dots) in the magnitude range $0.6 < J-K_s <1 $ and $K_s < 12$ (green dots). {\it Upper-right panel:} histogram showing the number of targets in the field (blue) and those observed (red) in the colour-magnitude domain selected to look for solar-like oscillating giants. The red part of the diagram ($J-K_s \gtrsim 0.6$) is populated mostly by giants. {\it Lower panels:} Same as in the upper panels, but for LRc01.}
         \label{fig:TGTlra01}
\end{figure}

We retrieved from EXODAT \citep{Deleuil2009} photometric information on all the stars in the field, as well as the targets observed. As shown in Fig. \ref{fig:TGTlra01}, targets within each field of view were selected largely on the basis of colour-magnitude criteria. 
Within the observed targets, solar-like oscillations were searched for in stars belonging to a limited colour-magnitude domain: $0.6 < J-K_s <1 $ and $K_s < 12$ \citep[see][]{Mosser2010}. Restricting to this domain, we find no significant difference in the target selection bias applied to LRc01 compared to LRa01. 

\subsection{Distributions of radius and mass}
Among the  435 (1626)  giants with detected solar-like oscillations in LRa01 (LRc01),  198 (678) have SNR high enough not only to detect \numax\ and \Dnu\ but also to have a detailed characterisation of the envelope corresponding to the oscillation power excess (see Sec. 2.2 in  \citep{Mosser2010}). In what follows we shall refer to these two samples as N2 and N3, following the notation introduced in \cite{Mosser2010}.
 
We now compare the distributions of radius and mass of stars in LRa01 and LRc01.
Since scaling relations were not tested at high luminosities we excluded from the sample the few stars with $L>200$ L$_\odot$, and stars with no 2MASS photometry available. The fraction of targets excluded represents $< 3\,\%$ of the whole sample of N3 data.

As described in \cite{Miglio2013a}, we applied the Kolmogorov-Smirnov (K-S) test to the distributions of mass and radius to quantify differences between the two populations. We considered both N2 and N3 sets of data, including stars of all radii, or restricting to giants with $R < 20$ R$_\odot$. When comparing radii, the null hypothesis (LRa01 and LRc01 samples  are drawn from the same parent distribution) cannot be rejected, or the difference between the two populations is marginally significant at best. On the other hand, we find the difference between the mass distribution of the two populations highly significant (K-S probability higher than 99.9\,\%), with stars in LRc01 having a lower average mass than those in LRa01.

To test the impact of statistical uncertainties on mass and radius on the K-S test, we perturbed the observed radii and masses adding a random offset  drawn from a Gaussian distribution having a standard deviation equal to the estimated uncertainty on the mass / radius.
We generated 1000 realisations of such distributions and performed a K-S test on each realisation. The distribution of results from the 1000 K-S tests confirms that while comparing radius  the null hypothesis cannot be rejected, the difference in the mass distribution is significant (in 95\,\% of the realisations the K-S probabilities are higher than 99\,\%) .

\section{Comparison with synthetic stellar populations}
\label{sec:synthetic}
For a meaningful comparison between observed and synthetic populations, we apply to the synthetic population the same selection criteria based on colour and magnitude adopted in the target selection. Moreover, as discussed in detail in \citep{Mosser2010}, since no significant bias in the distribution of the targets is present in the  \numax\, range between 6 $\mu$Hz and 80 $\mu$Hz,  we shall only consider observed and simulated stars with \numax\ in this frequency range. 
\begin{figure}
\centering
      \includegraphics[width=1\hsize]{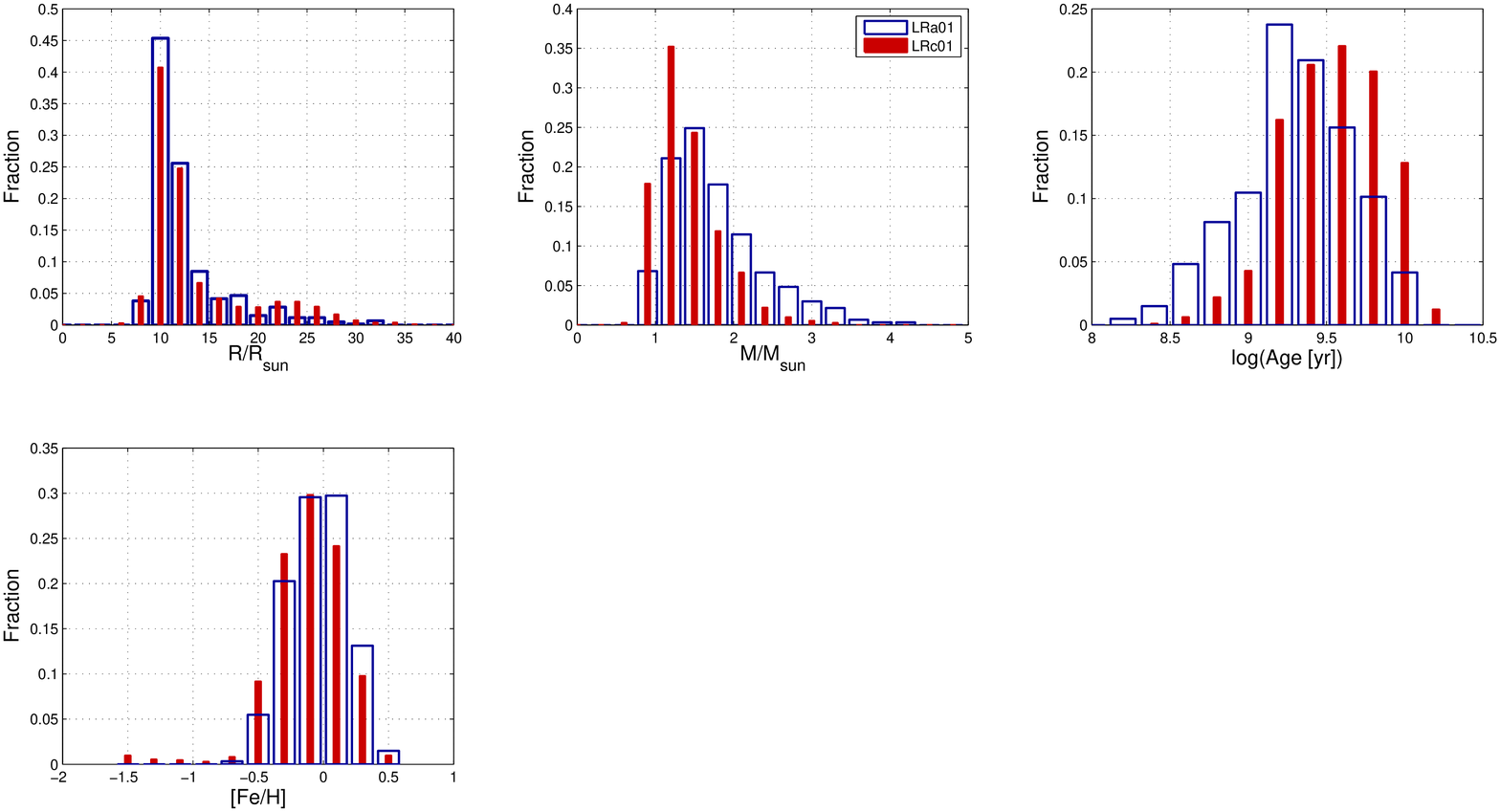}
\hspace*{0.4\hsize}
\begin{minipage}[]{0.55\hsize}
\vspace{-4.5cm}

      \caption{TRILEGAL simulations. Distributions of radius ({\it upper-left panel}), mass ({\it upper-middle panel}), age  ({\it upper-right panel}), and metallicity  ({\it lower-left panel}) of giants in synthetic populations computed with TRILEGAL for stars in the field LRa01 (blue) and LRc01 (red). Stars in the synthetic population were selected in magnitude ($K_s < 12$), colour ($0.6 < J-K_s <1 $), and \numax\, (6 $\mu$Hz $<$ \numax\, $<$ 80 $\mu$Hz) to account for target selection effects.}
         \label{fig:trile}
\end{minipage}

\end{figure}
\begin{figure}
\centering
      \includegraphics[width=.95\hsize]{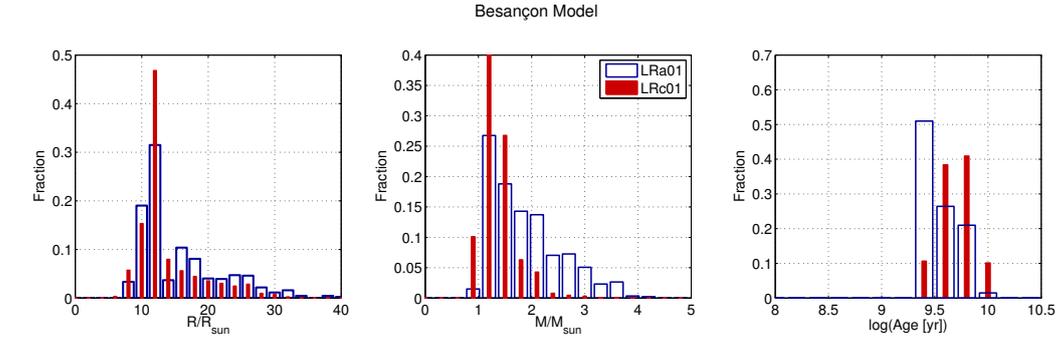}

      \caption{Same as Fig. \ref{fig:trile}, but for simulations with the Besan\c{c}on Model of the Milky Way. The age sampling of the Besan\c{c}on model is very coarse, resulting in an age resolution of only $\sim 1$ Gyr (see e.g. \citep{Robin2012})}.
         \label{fig:besa}
\end{figure}

We consider synthetic populations computed with two codes: the Besan\c{c}on Model of the Milky Way \citep[see e.g.][for a detailed description]{Robin03, Robin2012}, and with TRILEGAL \citep{Girardi05, Girardi2012}. TRILEGAL builds a geometric model for the Milky Way (MW), including main components such as the thin and thick
disks, halo and bulge, each one containing a particular stellar population. The latter
are defined by means of a star formation rate and age-metallicity relation. The main
geometric parameters of the MW components are then calibrated using wide-area
data for several lines-of-sight and for magnitudes at which the reddening is small
and at magnitudes for which problems such as photometric incompleteness and star/galaxy
separation are not an issue. Then, this simple model is applied everywhere in
the MW, for different filters, and at deeper magnitudes, making the implicit assumption
that the several MW components (but for the interstellar dust) are smoothly
distributed and quite uniform in their distributions of ages/metallicities.

Simulations with both TRILEGAL and  the Besan\c{c}on Model (see Fig. \ref{fig:trile} and \ref{fig:besa}) show that, although a similar distribution of radius is expected in LRa01 and LRc01, the age (hence mass) distribution of the two populations of giants is  expected to differ. Stars in LRc01 are expected to be older -on average- than those in LRa01.  

\begin{flushleft}
\begin{figure}
\begin{minipage}[t]{0.6\hsize}
      \includegraphics[width=\hsize]{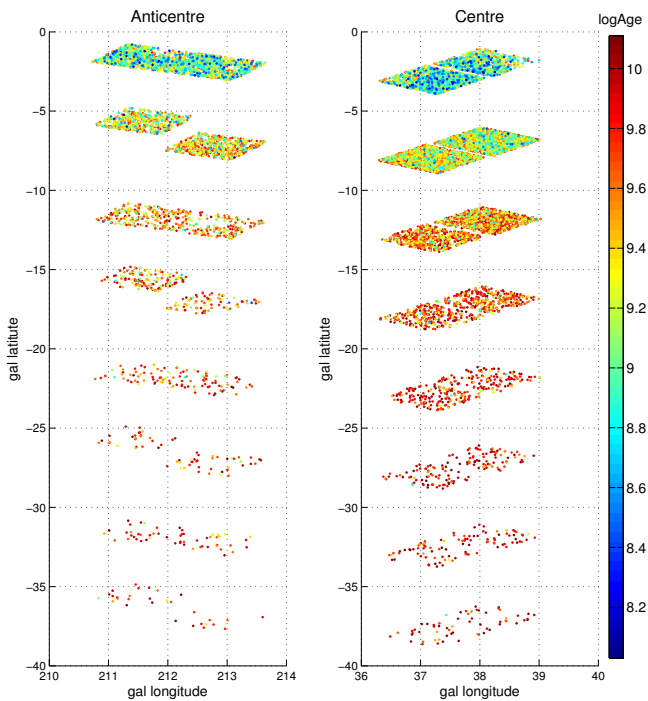}
      \end{minipage}
\hspace* {1cm}
\begin{minipage}[t]{0.3\hsize}
\vspace{-4cm}
     \caption{TRILEGAL: stars in simulated fields centred at the galactic longitude of  LRa01 ($l\sim 217^\circ$, {\it left panel}) and LRc01 ($l\sim 37^\circ$, {\it right panel}), and at decreasing galactic latitudes ({\it from top to bottom}). Stars are selected with the same colour-magnitude-$\nu_{\rm max}$ criteria as the observed targets (see Fig. \ref{fig:trile}), and their age is colour coded. }
         \label{fig:lat_long}
\end{minipage}
\end{figure}
\end{flushleft}

\begin{figure}
\begin{minipage}[t]{1.15\hsize}
\centering
\hspace*{-3.0cm}
      \includegraphics[width=.45\hsize]{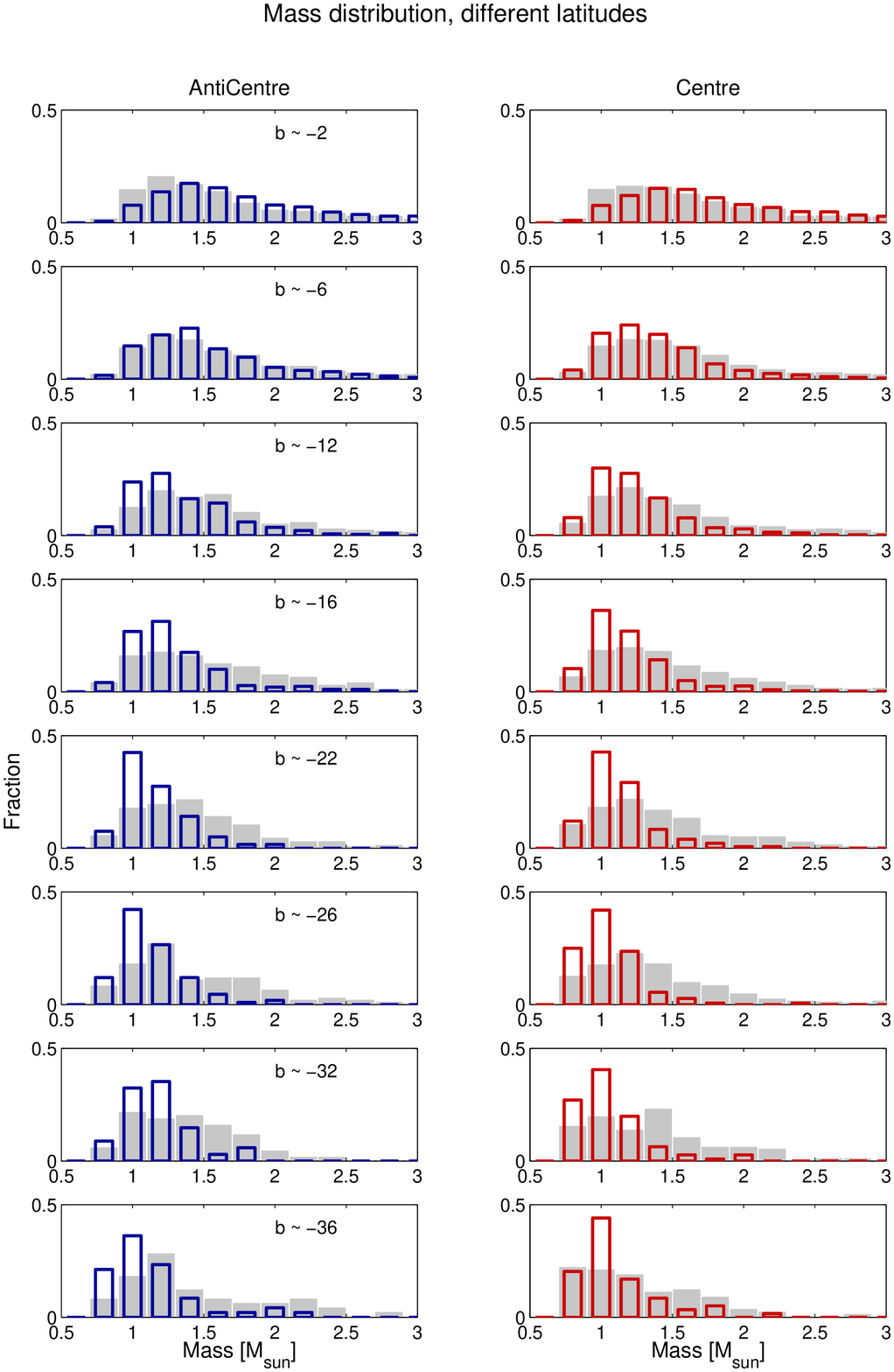}
   \includegraphics[width=.45\hsize]{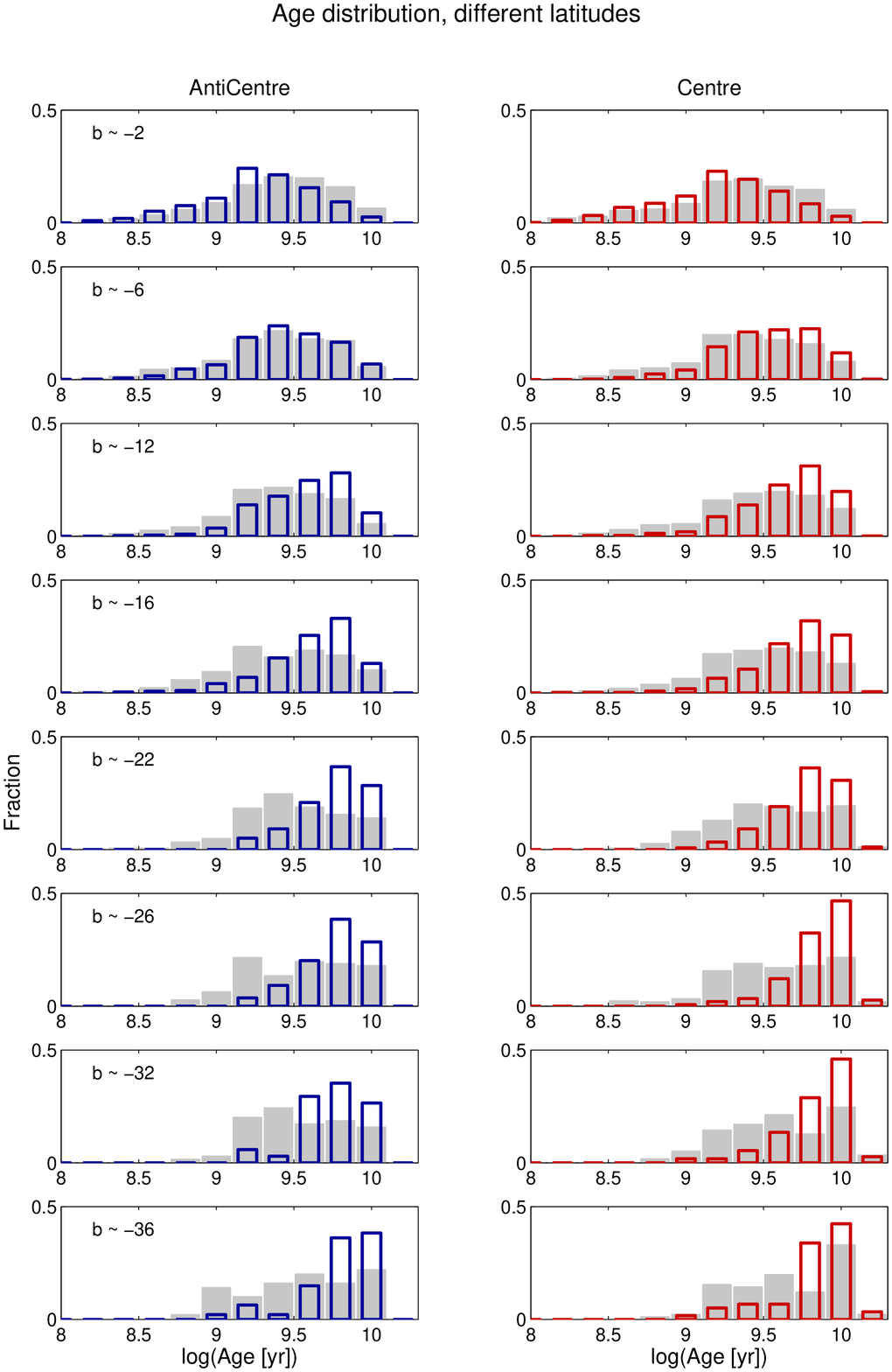}
\end{minipage}
     \caption{Open bars: same as Fig. \ref{fig:lat_long}, showing the mass distribution for each field at different latitudes and at the longitude of LRa01 ({\it left panel})  and LRc01 ({\it right panel}). For reference, solid grey bars show the distribution of mass in simulations with no age-dependent thin-disk vertical scale height. The galactic latitude of LRc01 and LRa01 is $b\simeq-7^\circ$ and $b\simeq-2^\circ$, respectively. Same as Fig. \ref{fig:lat_long}, showing the age distribution of  synthetic populations in fields of decreasing latitudes ({\it from top to bottom}) and of the longitude of LRa01 ({\it left panel})  and LRc01 ({\it right panel}), with (open bars) and without (grey bars) an age-dependent vertical density scale height for the thin disk.}
         \label{fig:mass_lat}
\end{figure}

To check whether in the synthetic populations the difference between LRc01 and LRa01 is due to a different latitude,  we simulated with TRILEGAL CoRoT fields at decreasing latitudes, and fixing the longitude to that of LRa01 and LRc01 (see Fig. \ref{fig:lat_long}). The expected distribution of mass and age are presented in Figure \ref{fig:mass_lat}.  In the magnitude range observed by CoRoT, while a broad spectrum of age / mass is expected for giants in  low-latitude fields, only old stars are expected to populate high-latitude fields. In these models the difference in the mass distribution of stars in LRc01 ($b \sim -7^\circ$) and LRa01 ($b \sim -2^\circ$)  is dominated by the different latitudes (hence different average height below the plane).

Our aim here is not to find the synthetic population that best matches the data, but to show that the difference we see in the observed distribution is in qualitative agreement with the simulations which include an increase of the disk scale height with age. To check the effect of this assumption on the predicted distributions of mass and age, we computed with TRILEGAL synthetic populations imposing an age-independent vertical scale height of the disk, and setting the latter to 300 pc.  As shown in Fig. \ref{fig:mass_lat} in this case we do not expect to see a clear mass / age gradient with latitude.

TRILEGAL simulations also suggest that the sample observed in the two fields is dominated by thin disk stars, with a combined fraction of thick disk and halo stars $\lesssim 10\,\%$. The bulge component, while obviously absent in LRa01, contributes by  $\lesssim 1\,\%$ to the population in LRc01.

\section{Grid based method: age estimates}
The age of  RGB and red-clump  stars is largely determined by their main-sequence lifetime and hence, to a first approximation, by their mass.  To estimate the age of giants in the two observed populations we use  PARAM, a Bayesian stellar parameter estimation method described in detail in  \citep{DaSilva2006}. PARAM is largely inspired by the method developed in
\citep{Joergensen2005}, which is designed to avoid statistical biases and to take error estimates of all observed quantities into consideration.
For this work we adapted PARAM to include as additional observational constraints \numax, \Dnu, and, when available, the evolutionary status of the star (core- Helium or hydrogen-shell burning phase), as determined by the period spacing of gravity-dominated modes (see \cite{Mosser2011}). Having knowledge of the evolutionary status is particularly useful in the age determination of stars with an estimated radius typical of RC giants. Since the mass of core-He burning stars is likely to be affected by mass loss, the age-mass relation in red-clump stars may be different from that of RGB stars (see e.g. Fig. 1 in \cite{Miglio2012}). This extended version of PARAM will  be made available, via an interactive web form, at the URL {\tt http://stev.oapd.inaf.it/param}.

The main assumptions of the method are: a maximum value for the age, which we set to 10 Gyr, a prior that all ages are equally likely and, given that no information on the metallicity of these stars is available, a broad 
 gaussian distribution is assumed as prior for the metallicity. The gaussian distribution is centred at [Fe/H]=-0.5, has a standard deviation $\sigma=1$, and ranges from [Fe/H] $= -2.3$ to [Fe/H] $= +0.3$. Finally, an implicit hypothesis of the method is that theoretical models provide a reliable description of the way stars of different mass, metallicity, and evolutionary stage distribute along the red-giant phase.

We apply this parameter estimation method to the sample of giants observed in LRc01 and LRa01. 
The uncertainty on the age as estimated using PARAM is of the order of 30-40\,\%. 
The results obtained with a model-based approach support the interpretation that the differences between the observed population in LRa01 and LRc01 are due to a different mass, hence age, distribution.  By performing K-S tests on 1000 realisations of the populations taking into account Gaussian uncertainties in the age determination, we find the difference between the age distributions of the two populations highly significant.

The precision of the age determination will significantly improve when constraints on the photospheric chemical composition will be available. In this context, and perhaps even more relevant, are tests to be performed on the accuracy of the age (and mass) estimates obtained assuming scaling relations for \numax\, and \Dnu\ (as discussed at the end of Sec. \ref{sec:test}). Making use of the full set of seismic constraints (individual frequencies of both acoustic and gravity-dominated modes) will possibly provide more stringent estimates of systematic uncertainties.  We consider this our next, high-priority, step in this line of work.


\begin{acknowledgement}
AM is thankful to the organisers of the conference for financial support.
TM acknowledges financial support from Belspo for contract PRODEX GAIA-DPAC. JM and MV acknowledge financial support from Belspo for contract PRODEX COROT.
\end{acknowledgement}

\bibliographystyle{epj}
\bibliography{andrea_m.bib}


\end{document}